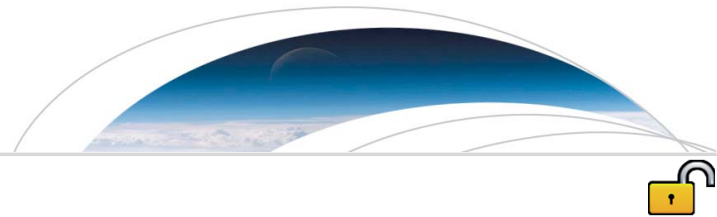

# Swarm in situ observations of *F* region polar cap patches created by cusp precipitation

L. V. Goodwin[1], B. Iserhienrhien[1], D. M. Miles[2], S. Patra[3], C. van der Meeren[4], S. C. Buchert[5], J. K. Burchill[6], L. B. N. Clausen[3], D. J. Knudsen[6], K. A. McWilliams[1], and J. Moen[3]

[1]Institute of Space and Atmospheric Studies, Department of Physics and Engineering Physics, University of Saskatchewan, Saskatoon, Saskatchewan, Canada, [2]Department of Physics, University of Alberta, Edmonton, Alberta, Canada, [3]Department of Physics, University of Oslo, Oslo, Norway, [4]Birkeland Centre for Space Science, Department of Physics and Technology, University of Bergen, Bergen, Norway, [5]Swedish Institute of Space Physics, Uppsala, Sweden, [6]Department of Physics and Astronomy, University of Calgary, Calgary, Alberta, Canada

**Abstract** High-resolution in situ measurements from the three Swarm spacecraft, in a string-of-pearls configuration, provide new insights about the combined role of flow channel events and particle impact ionization in creating *F* region electron density structures in the northern Scandinavian dayside cusp. We present a case of polar cap patch formation where a reconnection-driven low-density relative westward flow channel is eroding the dayside solar-ionized plasma but where particle impact ionization in the cusp dominates the initial plasma structuring. In the cusp, density features are observed which are twice as dense as the solar-ionized background. These features then follow the polar cap convection and become less structured and lower in amplitude. These are the first in situ observations tracking polar cap patch evolution from creation by plasma transport and enhancement by cusp precipitation, through entrainment in the polar cap flow and relaxation into smooth patches as they approach the nightside auroral oval.

## 1. Introduction

Polar cap patches are defined as 100 to 1000 km scale structures in the *F* region polar cap that have an enhanced plasma density at least twice that of the background. Polar cap patches are believed to be generated in the vicinity of the dayside cusp and are transported toward the nightside across the polar cap along the streamlines of convection [*Crowley*, 1996]. Such structures are associated with scintillations of transionospheric signals [*Basu et al.*, 1998; *Prikryl et al.*, 2011; *Moen et al.*, 2013], making them of particular interest to the space weather community and, for example, Global Navigation Satellite Systems (GNSS) [*Conker et al.*, 2003]. In the classical picture, high-density polar cap patches are islands of plasma cut off from the dayside tongue of ionization [*Lockwood and Carlson*, 1992; *Lockwood et al.*, 2005; *Carlson et al.*, 2004; *Moen et al.*, 2006] and brought into the polar cap near noon by the solar wind-driven polar cap convection [e.g., *Foster*, 1984; *Foster et al.*, 2005].

Low-density patches may be produced locally in the cusp by particle impact ionization [*Kelley et al.*, 1982; *Weber et al.*, 1984; *Walker et al.*, 1999; *Moen et al.*, 2012]. *Zhang et al.* [2013] presented a case where they suggested that low-density patches interspaced a sequence of high-density patches. Unfortunately, there is little in situ data available to distinguish various regimes of *F* region plasma structures or their sources. For example, *Smith et al.* [2000] made in situ observations of cusp density structures that they attributed to precipitation; however, they had to infer the existence of precipitation by the presence of high-frequency (HF) radar backscatter, making a definitive conclusion difficult.

As patches are transported across the polar cap, multiple processes and instabilities will act on and evolve electron density gradients present in the plasma. For example, the gradient-drift instability (GDI) is believed to operate efficiently on the trailing edge of electron density structures but not on the leading edge, contributing to the conventional asymmetric and flat-topped shape of a polar cap patch [e.g., *Tsunoda*, 1988; *Basu et al.*, 1994].

This letter presents novel high-resolution multipoint observations from the Swarm spacecraft, in an ideal noon-midnight orbit, providing new insights about the creation, transport, and evolution of polar cap patches. Here a case of polar cap patch formation is presented where a reconnection-driven low-density relative westward flow channel is eroding the dayside extreme ultra violet (EUV) plasma but where particle impact ionization in the cusp dominates the initial plasma structuring. These density structures are then transported across the polar cap along the streamlines of convection. Interestingly, smoother patch structures were observed deeper into the polar cap. These observations contrast the classical paradigm, where plasma structures would be expected to develop on the trailing edge of initially smooth plasma islands and then eat their way through [e.g., *Gondarenko and Guzdar*, 2004].

## 2. Instrumentation

This study uses in situ observations of the *F* region ionosphere from instruments on the Swarm spacecraft [*Friis-Christensen et al.*, 2008], orbiting at 500 km altitude. The Swarm mission consists of three spacecraft—Swarm A, Swarm B, and Swarm C—that, during postlaunch commissioning, were traveling along a common trajectory in a string-of-pearls configuration across the polar cap from magnetic noon to magnetic midnight. This study focuses on observations from 06 to 10 universal time (UT) on 30 December 2013. Swarm B led the constellation, Swarm A followed 59 s later, and Swarm C another 108 s behind Swarm A. The lateral separation of the three spacecraft ranged from 4 to 25 km over the region of interest. A detailed description of the Swarm data products is provided by Floberghagen et al. (this issue).

Electron density and temperature are measured at two samples per second by the Langmuir probe portion of the Electric Field Instrument (EFI) [*Friis-Christensen et al.*, 2008]. Ion flow velocity is measured at two samples per second by the Thermal Ion Imagers, which are also part of the EFI. Due to an uncertainty in the ion velocity baseline at this early stage of the Swarm mission, an offset was subtracted (roughly 800 m/s) from the cross-track flow component to achieve an approximate zero ion velocity equatorward of the cusp boundary. Therefore, this study only considers the short-term relative changes in ion velocity. Additionally, due to the roughly North-South orbit of the spacecraft during this period, the cross-track ion velocity will be interpreted as the eastward ion velocity.

The presence of field-aligned currents (FACs) is inferred from the deflections in the cross-track (east-west geographic) component of the Vector Field Magnetometer. The field is measured at 50 samples per second and downsampled to one sample per second. The CHAOS-4 geomagnetic field model [*Olsen et al.*, 2014], the MF7 seventh-generation lithospheric magnetic field model [*Maus et al.*, 2007], and the POMME8 model of the quasi-static magnetospheric currents [*Maus et al.*, 2010] have been subtracted from the measurements to estimate the residual magnetic field due to FACs.

Swarm observations are supported by all-sky imager (ASI) and coherent scatter radar observations, along with measurements of the interplanetary magnetic field (IMF). Red-line (630.0 nm) all-sky airglow data from the ASI in Longyearbyen, Svalbard (geomagnetic: 75.24°N, 111.21°E) [*Oksavik et al.*, 2012], were projected to 250 km altitude. The backscatter data used in this study were from the Super Dual Auroral Radar Network (SuperDARN) HF radar in Hankasalmi, Finland (geomagnetic: 59.1°N, 104.5°E) [*Lester et al.*, 2004; *Chisham et al.*, 2007]. Solar wind and IMF parameters were obtained from the OMNIweb database [*King and Papitashvili*, 2005].

## 3. Observations

Figure 1 shows the winter dayside cusp, as observed on 30 December 2013 during a quiet ($K_p = 0-1$) period. The IMF $B_z$ was consistently southward on the order of $-5$ nT, and the IMF $B_y$ was consistently positive on the order of $+2$ nT, both for more than 1 h prior to the first pass. This suggests active dayside reconnection with westward flow in the dayside ionosphere [e.g., *Ruohoniemi and Greenwald*, 2005].

Figures 1a and 1b show the trajectory of the third pass of the three spacecraft through the fields of view of the Longyearbyen ASI and the Hankasalmi SuperDARN HF radar, respectively. The Swarm spacecraft are in a near-polar orbit traveling through the region of interest in the Scandinavian sector and across the polar cap. At the location of Swarm B (red dot) shown in Figure 1, auroral emissions and signatures of poleward moving auroral forms recorded by the ASI are collocated with SuperDARN HF backscatter, which is characteristic of the cusp [*Moen et al.*, 2001]. A relative westward ion flow, indicated by vectors along the Swarm trajectory in Figure 1a, are observed in a region of auroral emissions within the cusp.

Figures 1c–1f show the electron/plasma density, eastward ion velocity deviation, electron temperature, and eastward magnetic field residual, respectively, from Swarm B. A notable density depletion appears in

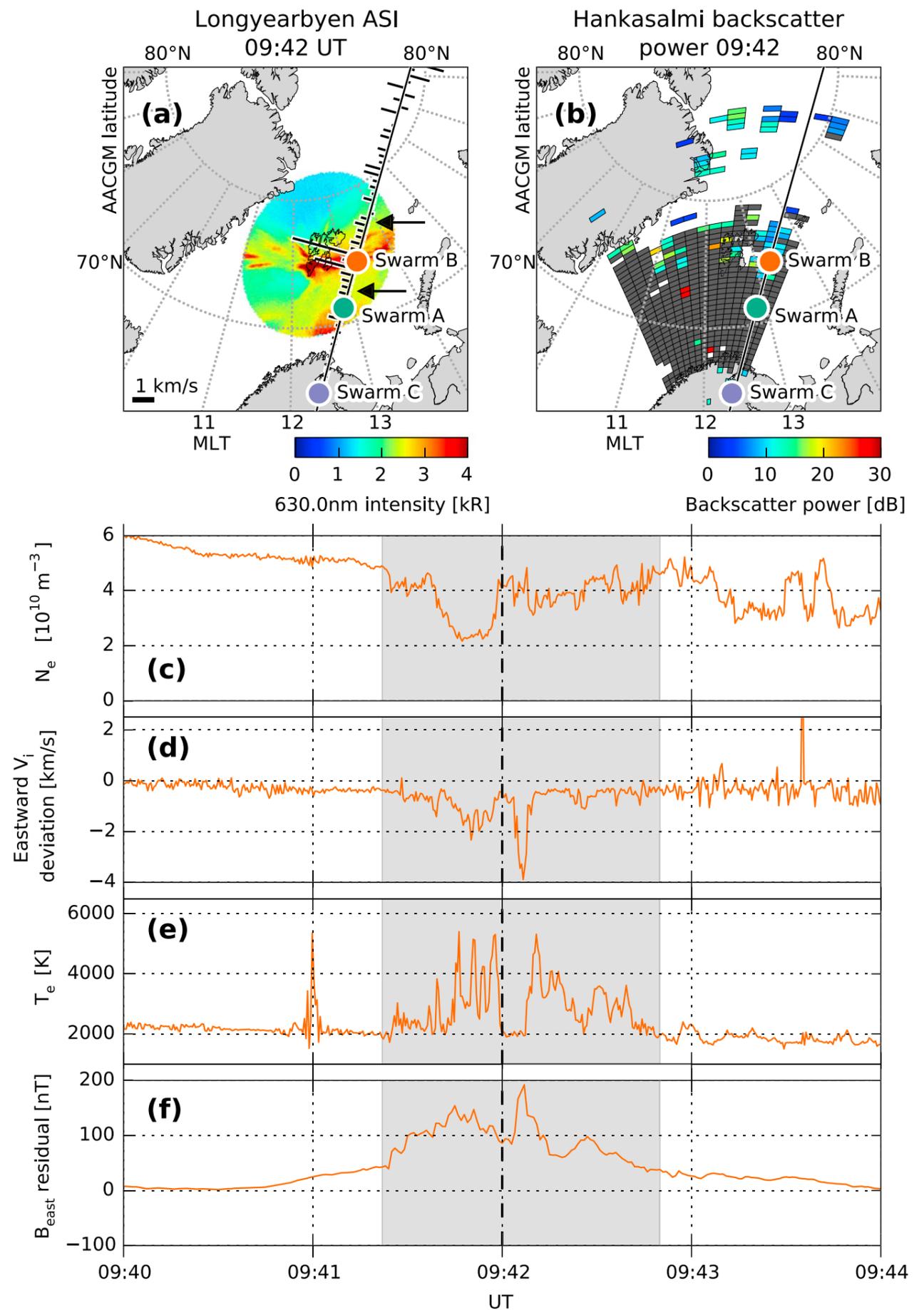

**Figure 1.** Observations from 30 December 2013. (a) Optical emissions at 630.0 nm from Longyearbyen ASI at 09:42 UT with Swarm trajectory and spacecraft locations. Velocity vectors along the track indicate the relative east-west (geographical) ion drift from Swarm B during the third pass. The black arrows indicate the extent of the shaded region in Figures 1c–1f. (b) Backscatter power at 09:42 UT from the SuperDARN Hankasalmi HF radar (ground scatter shown in grey). (c) Electron density ($N_e$). (d) Eastward ion velocity deviation ($V_i$). (e) Electron temperature ($T_e$). (f) Eastward magnetic field residual ($B_{east}$ residual). Data in Figures 1c–1f are taken from the third pass of Swarm B. The vertical dashed line indicates 09:42 UT.

Figure 1c just before 09:42 UT. This was coincident with an approximate 2 km/s westward deflection in the relative east-west ion flow component (Figure 1d), an increase in the electron temperature (Figure 1e), and an eastward (cross-track) magnetic perturbation suggesting an upward FAC (Figure 1f). The shaded area identifies the active cusp electron precipitation region, inferred from an increase in electron temperature, structured electron density, structured FACs, and auroral emissions. The open-closed boundary (OCB) is taken to be the equatorward edge of this shaded area (consistent with the signatures in *Carlson et al.* [2004]). The OCB lies within solar EUV-ionized plasma just equatorward of the density depletion. Poleward of the density depletion region, just after 09:42 UT, there are steep gradients in the east-west magnetic field perturbations superimposed on the general downward trend (Figure 1f). This suggests that smaller-scale FACs are superimposed on a larger-scale upward FAC.

Figures 2a–2c show the electron density profiles measured by the three spacecraft during three successive passes, along with the relative eastward ion velocity measured by Swarm B in the cusp precipitation region. The density profiles are plotted as a function of along-track distance from 60° Altitude-Adjusted Corrected Geomagnetic (AACGM) latitude [*Baker and Wing*, 1989]. This allows for an accurate comparison of the relative positions of the structures measured by the three spacecraft. In all of these passes, the Swarm spacecraft cross 75° magnetic latitude (MLAT) between 12 and 13 magnetic local time (MLT) on the dayside and between 21 and 22 MLT on the nightside, passing through the duskside polar cap. This means that each spacecraft pass has roughly the same trajectory with respect to MLAT and MLT. The spacecraft enter the cusp region at 06:30 UT (pass 1, Figure 2a), 08:05 UT (pass 2, Figure 2b), and 09:40 UT (pass 3, Figure 2c), respectively. The shaded areas mark the cusp precipitation region in each pass as inferred by the increased structuring in the electron temperature.

At the beginning of each pass in Figures 2a–2c, the three spacecraft encountered gradually decreasing electron density consistent with passing from dayside into nightside. In all three passes, all spacecraft observe a distinct depletion in electron density in the equatorward portion of the cusp. This depletion is approximately 150–200 km wide (along-track) and is encountered at a lower MLAT with each pass (~74.2 MLAT for pass 1, ~73.8 MLAT for pass 2, ~73.2 MLAT for pass 3). The minimum density in the depletion region is nearly constant for the nine spacecraft passes over a total of 3 h. This minimum density is independent of the photoionized electron density equatorward of the cusp, but it does match the baseline density observed deep in the polar cap (described further in section 4). Poleward of the depletion region, while still in the cusp precipitation region (i.e., the shaded region in Figure 1), all spacecraft encountered an increased and highly structured electron density during all passes.

## 4. Analysis and Discussion

Figure 2 shows a well-defined and consistent density depletion region in the equatorward portion of the cusp over a period of 3 h. Two processes can create plasma depletion: local recombination or transport from a lower-density region. *Valladares et al.* [1996] and *Pitout and Blelly* [2003] describe how an enhanced recombination rate leads to a depletion of plasma and an increase in ion temperature via ion frictional heating. In the EFI observations, however, there was no increase in the horizontal ion temperature (not shown). Conversely, *Lockwood et al.* [2005] and *Zhang et al.* [2011] explained that, in the case of positive IMF $B_y$, a density minimum results from low-density plasma transported westward into the cusp from the postnoon sector. The observations presented here show a distinct and prominent relative westward flow velocity enhancement as each spacecraft transits from the dayside EUV region into the density depletion region, a signature characteristic of magnetopause reconnection. This is consistent with the proposed plasma transport mechanism of *Lockwood et al.* [2005] and *Zhang et al.* [2011].

Figure 1 reveals the coincidence of the density depletion with the relative westward flow. From Figure 2, the density depletion region and the OCB are seen to be migrating equatorward between successive Swarm passes, consistent with the expansion of the polar cap when the IMF is southward. As the OCB migrates equatorward, the relative westward flow channel erodes regions of high-density plasma, mixing them with low-density plasma from the relative westward flow and incorporating them into the overall polar cap convection. Poleward of the flow channel, density structures are associated with the region of upward FAC (electron precipitation) consistent with *Lockwood et al.* [2005]. *Oksavik et al.* [2010] further pointed out that, due to the increased magnetic tension toward noon, a plasma packet may rotate 180° when entering the polar cap. *Rinne et al.* [2007] mapped the local flow disturbance of pulsed flux transfer events in the postnoon and found them to be consistent with such a rotation. In the Swarm observations, a relative westward flow channel is situated immediately poleward of the OCB consistent with it being driven by the IMF $B_y$ component of the magnetic tension force. The two-dimensional flow picture cannot be fully resolved by the current data set; however, it suggests that the electron density is depleted as the solar EUV-ionized plasma is initially pulled westward toward noon, and after the magnetic tension force has been released, it will turn clockwise into the polar cap with low-density polar cap plasma streaming in behind.

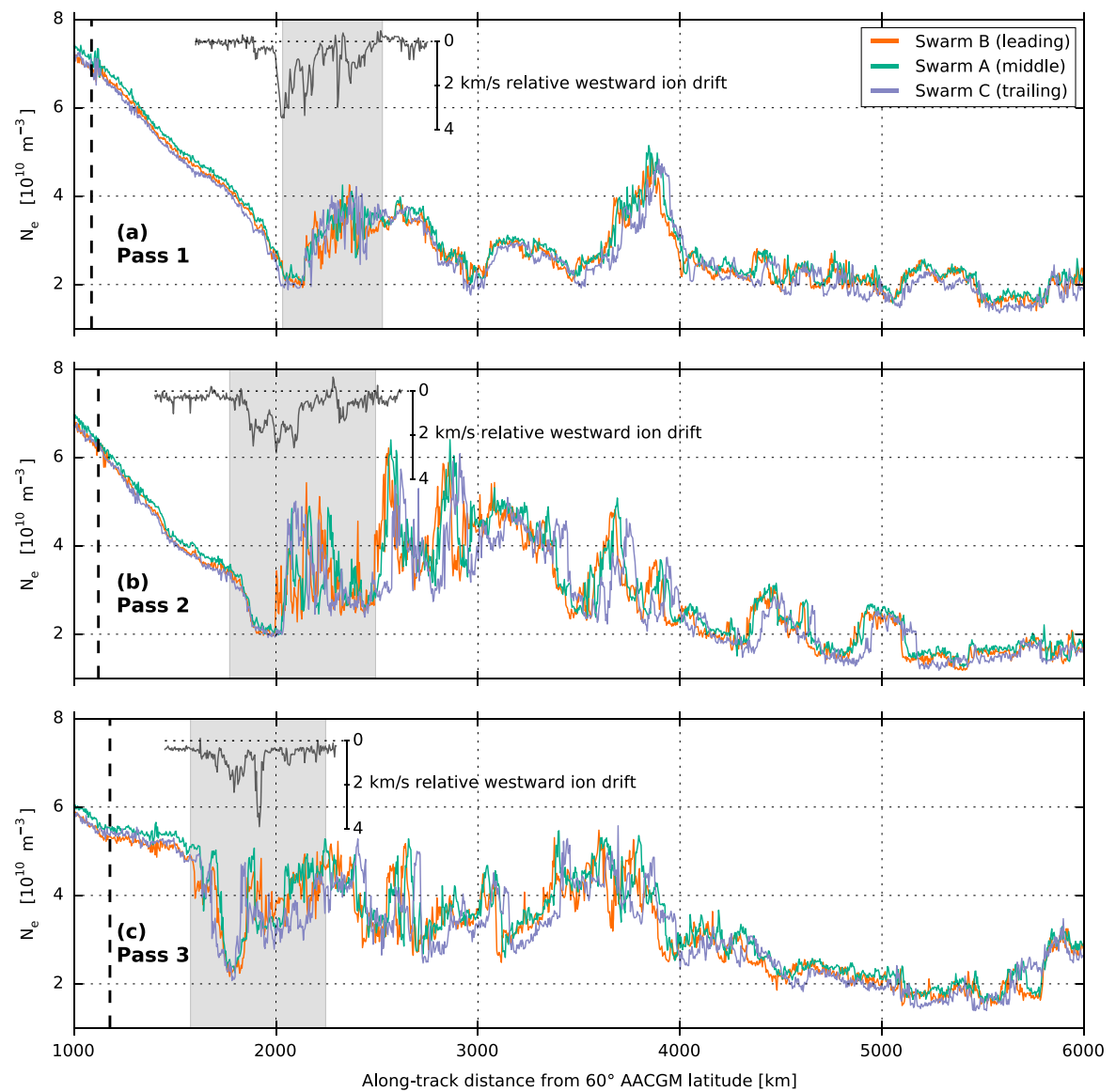

**Figure 2.** Observations from 30 December 2013. The measured density from all three spacecraft and all three passes is shown as a function of along-track distance, where the shaded areas indicate the cusp precipitation region in each pass, the black line plots indicate the corresponding relative eastward ion velocity from Swarm B, and the dashed line indicates when the Sun is 5° below the horizon. The consistent density depletion in the equatorward portion of the cusp moves between passes, irrespective of the dayside density. Note also that the density depletion is the same for all three spacecraft on all passes.

The shaded regions in Figure 3 identify eight enhanced density regions from Figure 2b, which will be referred to as "features" I–VIII. Since density is plotted against along-track distance from a set geomagnetic latitude, stationary structures are expected to be observed at the same location by the three different spacecraft. Conversely, structures with an along-track velocity should show a separation where each successive spacecraft (Swarm B, Swarm A, then Swarm C) encounters the structure farther along track. Features II through VIII do show along-track motion consistent with traveling parallel to the streamlines of convection. In contrast, feature I, located in the cusp precipitation region, does not show the same prominent motion. This is interpreted as the mixture of transport and particle impact ionization structuring masking the motion.

When the spacecraft cross the polar cap, the along-track motion of the density structures creates a time history of plasma evolution. This time history shows the production of plasma structures due to cusp precipitation and the evolution and transport of plasma into and across the polar cap. A comparison of the successive electron profiles in Figure 3 reveals an overall along-track motion that becomes clear for the less structured features seen farther along track. This time history also shows that electron density features that are closer to the region of particle impact ionization are more structured than those farther along track, which are evolving into the classic asymmetric and flat-topped shape of a polar cap patch. The time history of patch morphology created by this technique provides more context and can resolve shorter timescale changes than previous in situ observations [e.g., *Smith et al.*, 2000].

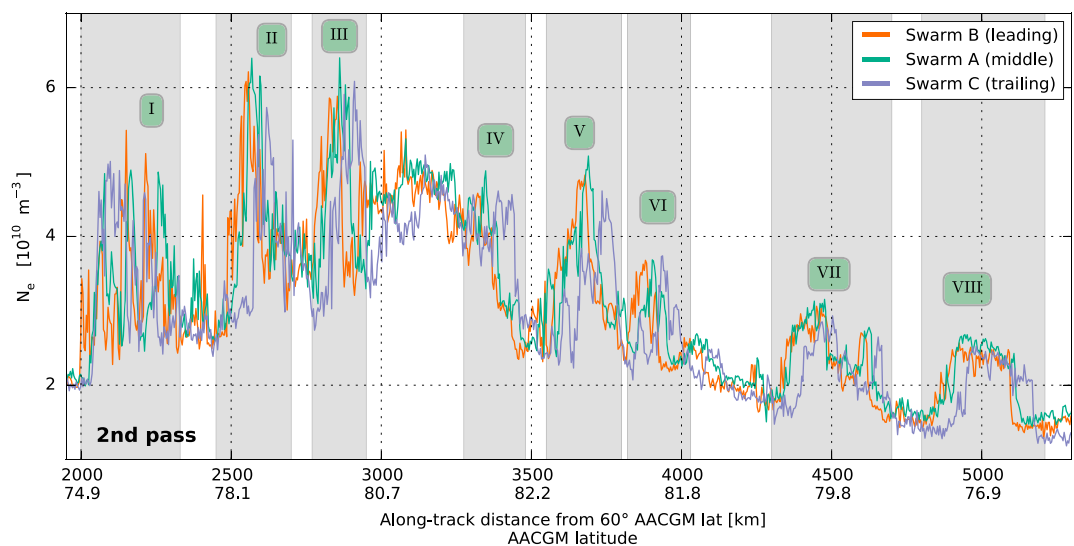

**Figure 3.** Observations from 30 December 2013. A magnified view of Figure 2b showing the along-track (poleward) motion of density structures. Features encountered in sequence by Swarm B (red), Swarm A (green), and Swarm C (blue) indicate along-track (poleward) motion. Shaded regions I–VIII identify structures discussed in the text.

Table 1 summarizes quantitatively the patch evolution seen in Figure 3, as well as the feature evolution seen in passes 1 and 3. The estimated along-track velocity for the different features (calculated from the change in along-track location at which a particular feature is encountered by the three spacecraft in one pass; see Table 1, column 2) are generally in the range of 200–500 m/s and are consistent with typical convection velocities in the polar cap (as seen in *Pettigrew et al.* [2010, and references therein]). In passes 1 and 2 the feature thickness (the approximate along-track distance in which a density feature is enhanced above the immediately surrounding background; see Table 1, column 3) generally increases with along-track distance. In all three passes, the feature thickness ranges from roughly 50 km to 300 km, making these observations consistent with past studies of polar cap patch thickness [e.g., *Hosokawa et al.*, 2014]. These multipoint in situ Swarm observations can further be used to study processes such as the GDI, which may be acting on the newly created electron structures (e.g., Spicher et al., this issue). This is, however, beyond the scope of this article.

**Table 1.** Along-Track Velocity and the Thickness for the Eight Features (I Through VIII) Identified in Figure 3, as Well as Consecutive Features Seen in Passes 1 and 3[a]

| Feature | Along-Track Velocity (m/s) | Feature Thickness (km) |
|---|---|---|
| *Pass 2 (Featured in Figure 3)* | | |
| I | - | 40–60 |
| II | 180–230 | 80–140 |
| III | 400–450 | 140–160 |
| IV | 300–530 | - |
| V | 300–500 | 150–180 |
| VI | 300–420 | 90–120 |
| VII | 180–350 | 300–400 |
| VIII | 300–530 | 270–320 |
| *Pass 1* | | |
| I | - | 200–250 |
| II | - | 190–210 |
| III | 60–140 | 340–360 |
| IV | 90–180 | 310–320 |
| V | 160–180 | 80–100 |
| VI | 90–270 | 110–130 |
| VII | 120–320 | 230–250 |
| *Pass 3* | | |
| I | - | 50–140 |
| II | - | 250–360 |
| III | 160–180 | 130–150 |
| IV | 180–470 | 100–130 |
| V | 100–470 | 70–110 |
| VI | 240–680 | 170–240 |
| VII | 140–460 | 400–420 |

[a]The given ranges represent the variability when comparing different pairs of spacecraft (for example, comparing Swarm B to A and Swarm A to C).

**Acknowledgments**
This study makes use of data from the Swarm spacecraft mission which is funded and managed by the European Space Agency. Swarm data can be accessed online at: http://earth.esa.int/swarm. All EFI data used in this study were produced by the Level 1b prototype data processor. Clausen and the ASI at Longyearbyen are supported by the Norwegian Research Council under contract 230935. The ASI data are available at http://tid.uio.no/plasma/aurora. The Hankasalmi SuperDARN radar is funded by the Radio and Space Plasma Physics Group at the University of Leicester and is available at http://vt.superdarn.org. The authors thank Mark Lester for use of the data. The OMNI data were obtained from the GSFC/SPDF OMNIWeb interface at http://omniweb.gsfc.nasa.gov. Magnetic field residuals were calculated by Jan Rauberg and Claudia Stolle from GFZ Potsdam based on Swarm data. This study is a product of a PhD school arranged at Andøya Space Center in summer 2014. This was an activity within the Canada-Norway Rocket Science Training and Educational Program (CaNoRock STEP). Thanks to the University of Alberta, University of Bergen, University of Calgary, University of Oslo, University of Saskatchewan, University of Tromsø, and the University Centre in Svalbard for arranging this program. Canada-Norway Mobility funds for the CaNoRock STEP were provided by the Norwegian Centre for International Cooperation in Education (SiU), project NNA-2012/10099. This project has also received economic support from the Research Council of Norway, grant 230996. Miles is supported by a Canadian NSERC PGSD2 graduate scholarship. Goodwin and Iserhienrhien are supported through an NSERC Discovery Grant given to J-P St-Maurice. Burchill is supported financially by the Canadian Space Agency. The authors would like to thank Rune Floberghagen for his advice and guidance.

The Editor thanks Keisuke Hosokawa for his assistance in evaluating this paper.

In the region of features I, II, and III in Figure 3, the density floor increases from $2 \times 10^{10}$ m$^{-3}$ to $3 \times 10^{10}$ m$^{-3}$, on top of which the structured density maxima increase from $5 \times 10^{10}$ m$^{-3}$ to more than $6 \times 10^{10}$ m$^{-3}$. This indicates that particle impact ionization can be a dominant process in producing high-density structures on top of either EUV solar-ionized plasma that has been transported into the polar cap or the background density of the polar cap. Between features IV and V in Figure 3 the spacecraft encountered a low-density region of roughly $2 \times 10^{10}$ m$^{-3}$, which is seen several times throughout the along-track motion, including within the flow channel. This is likely the baseline density of the dark *F* region polar cap.

At the one extreme, density structures can be made by cusp particle precipitation. In the other extreme, the main density gradient can be produced by the intake of sunlit dayside plasma by transient dayside reconnection. The observed discrete electron temperature enhancements in the shaded region of Figure 1e (the cusp) are consistent with particle impact ionization initializing 10–100 km scale structures, as observed by *Kelley et al.* [1982] and *Moen et al.* [2012]. The highly structured plasma features evolve into lower-density, less structured polar cap patches as they drift toward the nightside. Furthermore, the more mature patches show the characteristic GDI-driven shape with a sharp leading edge, relatively uniform central region, and a structured and tapered trailing edge (features VII and VIII in Figure 3).

## 5. Conclusions

The Swarm string-of-pearls configuration provides a novel method to resolve, in situ, the creation, transport, and morphology of cusp electron density structures which become polar cap patches. In the presented observations from 30 December 2013 the spacecraft travel parallel to the inferred polar cap convection streamlines sampling a sequence of polar cap patches from the production region in the cusp through to the nightside.

1. A reconnection-driven relative westward flow channel and associated density depletion region separates the EUV plasma from a region of active electron precipitation in the cusp. The initial high-density structuring observed poleward of the density depletion in the cusp cannot be created by the direct injection of day-lit EUV plasma from lower latitudes.
2. The initial high-density plasma structures poleward of the westward flow channel appear to be the result of particle impact ionization that is superimposed on either solar EUV-ionized plasma or the baseline density of the polar cap. These structures are colocated with in situ magnetic signatures of FACs and auroral features observed in ground-based all-sky images.
3. The newly created and highly structured plasma evolves into lower-density, less structured polar cap patches as they transit the polar cap. These are the first in situ observations of a series of patches, entrained in the polar cap flow, traveling from their source in the cusp to the nightside auroral oval.
4. The observations contrast the classical view of polar cap patch formation that solar EUV-ionized patches are chopped off from the tongue of ionization and structure is added by the GDI acting on the trailing edge [*Gondarenko and Guzdar*, 2004]. Instead, we observe cusp precipitation that produces large plasma density structures [e.g., *Kelley et al.*, 1982; *Walker et al.*, 1999; *Smith et al.*, 2000] near noon that convect across the polar cap and smooth into classical polar cap patches. We conclude that the resulting kilometer-scale structures are the product of particle impact ionization in the cusp. The GDI process is potentially effective on the trailing edges of these gradients and may play an important role in the smoothing.

Multipoint sequential Swarm measurements, such as those presented, could be used for future quantitative studies of the processes thought to control polar cap patch dynamics. These data have the potential to provide more realistic initialization conditions in patch simulations, allowing for better predictions of patch irregularities and forecasting of space weather GNSS scintillation effects.

## References

Baker, K. B., and S. Wing (1989), A new magnetic coordinate system for conjugate studies at high latitudes, *J. Geophys. Res.*, *94*(A7), 9139–9143.
Basu, S., S. Basu, P. Chaturvedi, and C. Bryant (1994), Irregularity structures in the cusp/cleft and polar cap regions, *Radio Sci.*, *29*(1), 195–207.
Basu, S., E. Weber, T. Bullett, M. Keskinen, E. MacKenzie, P. Doherty, R. Sheehan, H. Kuenzler, P. Ning, and J. Bongiolatti (1998), Characteristics of plasma structuring in the cusp/cleft region at Svalbard, *Radio Sci.*, *33*(6), 1885–1899.
Carlson, H. C., K. Oksavik, J. Moen, and T. Pedersen (2004), Ionospheric patch formation: Direct measurements of the origin of a polar cap patch, *Geophys. Res. Lett.*, *31*, L08806, doi:10.1029/2003GL018166.

Chisham, G., et al. (2007), A decade of the Super Dual Auroral Radar Network (SuperDARN): Scientific achievements, new techniques and future directions, *Surv. Geophys.*, *28*(1), 33–109.

Conker, R. S., M. B. El-Arini, C. J. Hegarty, and T. Hsiao (2003), Modeling the effects of ionospheric scintillation on GPS/satellite-based augmentation system availability, *Radio Sci.*, *38*(1), 1001, doi:10.1029/2000RS002604.

Crowley, G. (1996), Critical review of ionospheric patches and blobs, in *Review of Radio Science 1993–1996*, edited by W. R. Stone, pp. 619–648, Oxford Univ. Press, New York.

Foster, J., et al. (2005), Multiradar observations of the polar tongue of ionization, *J. Geophys. Res.*, *110*, A09S31, doi:10.1029/2004JA010928.

Foster, J. C. (1984), Ionospheric signatures of magnetospheric convection, *J. Geophys. Res.*, *89*(A2), 855–865.

Friis-Christensen, E., H. Lühr, D. Knudsen, and R. Haagmans (2008), Swarm—An Earth observation mission investigating geospace, *Adv. Space Res.*, *41*(1), 210–216.

Gondarenko, N., and P. Guzdar (2004), Plasma patch structuring by the nonlinear evolution of the gradient drift instability in the high-latitude ionosphere, *J. Geophys. Res.*, *109*, A09301, doi:10.1029/2004JA010504.

Hosokawa, K., S. Taguchi, K. Shiokawa, Y. Otsuka, Y. Ogawa, and M. Nicolls (2014), Global imaging of polar cap patches with dual airglow imagers, *Geophys. Res. Lett.*, *41*, 1–6, doi:10.1002/2013GL058748.

Kelley, M. C., J. F. Vickrey, C. W. Carlson, and R. Torbert (1982), On the origin and spatial extent of high-latitude $F$ region irregularities, *J. Geophys. Res.*, *87*(A6), 4469–4475.

King, J., and N. Papitashvili (2005), Solar wind spatial scales in and comparisons of hourly Wind and ACE plasma and magnetic field data, *J. Geophys. Res.*, *110*, A02104, doi:10.1029/2004JA010649.

Lester, M., et al. (2004), Stereo CUTLASS—A new capability for the SuperDARN HF radars, *Ann. Geophys.*, *22*(2), 459–473.

Lockwood, M., and H. Carlson (1992), Production of polar cap electron density patches by transient magnetopause reconnection, *Geophys. Res. Lett.*, *19*(17), 1731–1734.

Lockwood, M., J. Moen, A. Van Eyken, J. Davies, K. Oksavik, and I. McCrea (2005), Motion of the dayside polar cap boundary during substorm cycles: I. Observations of pulses in the magnetopause reconnection rate, *Ann. Geophys.*, *23*(11), 3495–3511.

Maus, S., H. Lühr, M. Rother, K. Hemant, G. Balasis, P. Ritter, and C. Stolle (2007), Fifth-generation lithospheric magnetic field model from CHAMP satellite measurements, *Geochem. Geophys. Geosyst.*, *8*, Q05013, doi:10.1029/2006GC001521.

Maus, S., C. Manoj, J. Rauberg, I. Michaelis, and H. Lühr (2010), NOAA/NGDC candidate models for the 11th generation International Geomagnetic Reference Field and the concurrent release of the 6th generation Pomme magnetic model, *Earth Planets Space*, *62*(10), 729–735.

Moen, J., H. Carlson, S. Milan, N. Shumilov, B. Lybekk, P. Sandholt, and M. Lester (2001), On the collocation between dayside auroral activity and coherent HF radar backscatter, *Ann. Geophys.*, *18*(12), 1531–1549.

Moen, J., H. Carlson, K. Oksavik, C. Nielsen, S. Pryse, H. Middleton, I. McCrea, and P. Gallop (2006), EISCAT observations of plasma patches at sub-auroral cusp latitudes, *Ann. Geophys.*, *24*(9), 2363–2374.

Moen, J., K. Oksavik, T. Abe, M. Lester, Y. Saito, T. Bekkeng, and K. Jacobsen (2012), First in-situ measurements of HF radar echoing targets, *Geophys. Res. Lett.*, *39*, L07104, doi:10.1029/2012GL051407.

Moen, J., K. Oksavik, L. Alfonsi, Y. Daabakk, V. Romano, and L. Spogli (2013), Space weather challenges of the polar cap ionosphere, *J. Space Weather Space Clim.*, *3*, A02.

Oksavik, K., V. L. Barth, J. Moen, and M. Lester (2010), On the entry and transit of high-density plasma across the polar cap, *J. Geophys. Res.*, *115*, A12308, doi:10.1029/2010JA015817.

Oksavik, K., J. Moen, M. Lester, T. Bekkeng, and J. Bekkeng (2012), In situ measurements of plasma irregularity growth in the cusp ionosphere, *J. Geophys. Res.*, *117*, A11301, doi:10.1029/2012JA01783.

Olsen, N., H. Lühr, C. C. Finlay, T. J. Sabaka, I. Michaelis, J. Rauberg, and L. Tøffner-Clausen (2014), The CHAOS-4 geomagnetic field model, *Geophys. J. Int.*, *197*(2), 815–827.

Pettigrew, E., S. Shepherd, and J. Ruohoniemi (2010), Climatological patterns of high-latitude convection in the Northern and Southern Hemispheres: Dipole tilt dependencies and interhemispheric comparisons, *J. Geophys. Res.*, *115*, A07305, doi:10.1029/2009JA014956.

Pitout, F., and P.-L. Blelly (2003), Electron density in the cusp ionosphere: Increase or depletion?, *Geophys. Res. Lett.*, *30*(14), 1726, doi:10.1029/2003GL017151.

Prikryl, P., P. Jayachandran, S. Mushini, and R. Chadwick (2011), Climatology of GPS phase scintillation and HF radar backscatter for the high-latitude ionosphere under solar minimum conditions, *Ann. Geophys.*, *29*(2), 377–392.

Rinne, Y., J. Moen, K. Oksavik, and H. Carlson (2007), Reversed flow events in the winter cusp ionosphere observed by the European Incoherent Scatter (EISCAT) Svalbard radar, *J. Geophys. Res.*, *112*, A10313, doi:10.1029/2007JA012366.

Ruohoniemi, J., and R. Greenwald (2005), Dependencies of high-latitude plasma convection: Consideration of interplanetary magnetic field, seasonal, and universal time factors in statistical patterns, *J. Geophys. Res.*, *110*, A09204, doi:10.1029/2004JA010815.

Smith, A., S. Pryse, and L. Kersley (2000), Polar patches observed by ESR and their possible origin in the cusp region, *Ann. Geophys.*, *18*(9), 1043–1053.

Tsunoda, R. T. (1988), High-latitude $F$ region irregularities: A review and synthesis, *Rev. Geophys.*, *26*(4), 719–760.

Valladares, C., D. Decker, R. Sheehan, and D. Anderson (1996), Modeling the formation of polar cap patches using large plasma flows, *Radio Sci.*, *31*(3), 573–593.

Walker, I., J. Moen, L. Kersley, and D. Lorentzen (1999), On the possible role of cusp/cleft precipitation in the formation of polar-cap patches, *Ann. Geophys.*, *17*(10), 1298–1305.

Weber, E. J., J. Buchau, J. Moore, J. Sharber, R. Livingston, J. Winningham, and B. Reinisch (1984), $F$ layer ionization patches in the polar cap, *J. Geophys. Res.*, *89*(A3), 1683–1694.

Zhang, Q.-H., et al. (2011), On the importance of interplanetary magnetic field $|B_y|$ on polar cap patch formation, *J. Geophys. Res.*, *116*, A05308, doi:10.1029/2010JA016287.

Zhang, Q.-H., B.-C. Zhang, J. Moen, M. Lockwood, I. McCrea, H.-G. Yang, H.-Q. Hu, R.-Y. Liu, S.-R. Zhang, and M. Lester (2013), Polar cap patch segmentation of the tongue of ionization in the morning convection cell, *Geophys. Res. Lett.*, *40*, 2918–2922, doi:10.1002/grl.50616.